\documentclass[10pt,conference]{IEEEtran}
\IEEEoverridecommandlockouts
\usepackage{cite}
\usepackage{amsmath,amssymb,amsfonts}
\usepackage{algorithmic}
\usepackage{graphicx}
\usepackage{textcomp}
\usepackage{xcolor}
\usepackage{makecell}
\usepackage[frak=esstix]{mathalpha}
\usepackage{comment}
\usepackage{multirow}

\usepackage{commath}

\usepackage{pgfplots}
\pgfplotsset{
    compat=1.18,
    every axis/.append style={
        font=\footnotesize 
    }
}

\def\BibTeX{{\rm B\kern-.05em{\sc i\kern-.025em b}\kern-.08em
    T\kern-.1667em\lower.7ex\hbox{E}\kern-.125emX}}
\begin{document}

\title{ETSI ISG MAT: Bridging Multiple Access Techniques Research and 6G Standardisation\\

\thanks{This work was supported in part by WAVE-XL (PID2024-160457OB-I00) funded by MCIN/AEI/ 10.13039/501100011033, by “ERDF A way of making Europe.”, and European Commission’s Horizon Europe, SNS JU, research and innovation program grant agreement number 101139282, 6G-SENSES project.  
}
}

\author{\IEEEauthorblockN{Dheeraj Raja Kumar$^1$, Mert Ozates$^2$, Yong Jin Daniel Kim$^3$, David Vargas$^4$, Onur Dizdar$^5$, \\
Fehmi Emre Kadan$^5$, Carles Ant\'on-Haro$^1$, Sebastian Wagner$^6$, Bruno Clerckx$^7$, Malte Schellmann$^8$, \\
Xavier Mestre$^1$, H. M. Furqan$^9$, Kenza Abela$^9$, Shaima Abidrabbu$^9$}
\IEEEauthorblockA{$^1$Centre Tecnologic de Telecomunicacions de Catalunya (CTTC/iCERCA), Spain, \\
$^2$IHP- Leibniz Institute for High Performance Microelectronics, Germany, $^3$ Rose-Hulman Institute of Technology, USA, \\
$^4$BBC Research and Development, U.K., $^5$VIAVI Solutions Inc., U.K.,
$^6$EURECOM, Sophia-Antipolis, France, \\ $^7$Imperial College London, U.K., $^8$Huawei Technologies Munich, Germany, 
$^9$Vestel Electronics, Türkiye }
}
\maketitle

\begin{abstract}
Improved spectral efficiency and inter-user interference mitigation are important aspects of IMT-2030 and the ongoing 3GPP 6G study. ETSI established the Industry Specification Group on Multiple Access Techniques (ISG MAT) as a research and pre-standardisation activity to build wider consensus on downlink MAT for 3GPP-based 6G systems. Its first report, ETSI GR MAT 001 V1.1.1, provides a standards-oriented comparison of 3GPP-specified techniques, including OMA, MU-MIMO, and MUST, with candidate techniques comprising power-domain NOMA, RSMA, and cache-aided MU-MIMO. The report identifies operating conditions in which candidate MAT can improve spectral efficiency and assesses implications for transceiver processing, network assistance information, and reference-signal requirements. Ongoing work includes realistic 5G NR link-level evaluations and the study of MAT for non-terrestrial networks, providing timely technical evidence for 3GPP 6G standardisation discussions.
\end{abstract}


\section{Introduction and 6G Standardisation Context}
International Mobile Telecommunications for 2030 and beyond (IMT-2030) aims to address the diverse communication needs of humans and machines from 2030 onwards. It extends IMT-2020 with evolved capabilities and new service scenarios covering a wide range of performance requirements \cite{itu23, itu22}. Improved spectral efficiency is a key IMT-2030 capability \cite{itu23}, and more efficient use of finite spectrum resources will be needed to support increasing traffic volumes, heterogeneous services, and reliable performance across diverse deployment environments. Mobile network operators have also identified spectral efficiency as a factor in reducing the operational expenditure of future networks \cite{vab25}. Multiple Access Techniques (MAT), including Multi-User Multiple-Input Multiple-Output (MU-MIMO) and inter-user interference mitigation techniques, are candidate technologies for enhancing the IMT-2030 radio interface \cite{itu23, itu22}. Their potential benefits must be assessed alongside system aspects such as receiver complexity, reference-signal requirements, and network assistance information.

The 3GPP Release 20 Study on 6G Radio runs from the third quarter of 2025 to June 2027, while Release 21 will develop the first normative 6G specifications from March 2027 to the final ASN.1/OpenAPI freeze in March 2029 \cite{3gpp25, 3gpp26}. The timing of the 3GPP 6G study makes it important to evaluate candidate MAT against specified 3GPP features and assumptions. This paper presents the scope and activities of ETSI ISG MAT, examines its standards-oriented assessment of specified and candidate MAT in ETSI GR MAT 001 V1.1.1 \cite{mat26}, and discusses its ongoing work and relevance to the 3GPP 6G study.

\section{ETSI ISG MAT and GR MAT 001}
ETSI established the Industry Specification Group on Multiple Access Techniques (ISG MAT) at the end of 2024 as a research and pre-standardisation activity to build wider consensus on MAT for 6G systems based on 3GPP technical specifications. Its scope covers downlink MAT for the physical layer of the 3GPP radio interface, with the objective of improving transmission efficiency, including spectral efficiency, power consumption, latency, and user fairness. Open to ETSI and non-ETSI members, the group brings together participants from industry, academia, research organisations, and government administrations. ISG MAT produces informative deliverables in the form of Group Reports (GRs) that can be made available to 3GPP and other relevant fora for consideration in their 6G standardisation activities. This structure provides an open setting that aims at closing the gap between research and standardisation \cite{mat26}.

The first published report, ETSI GR MAT 001 V1.1.1, entitled \textit{Classification of Candidate Multiple Access Techniques for 6G and their Comparison with Specified 3GPP Features}, studies downlink techniques intended to improve spectral efficiency in the presence of inter-user interference \cite{mat26}. By using a common, standards-oriented framework, the report compares candidate MAT with baseline schemes specified by 3GPP, namely Orthogonal Multiple Access (OMA), MU-MIMO, and Multi-User Superposition Transmission (MUST). The candidate techniques are power-domain Non-Orthogonal Multiple Access (NOMA), Rate-Splitting Multiple Access (RSMA), and Cache-Aided MU-MIMO (CA MU-MIMO). The comparison takes into account transmit-processing architectures, receiver types, network assistance information provided to User Equipment (UE), Demodulation Reference Signal (DM-RS) requirements, and spectral-efficiency performance. By assessing these aspects from a radio-interface implementation and standardisation perspective, GR MAT 001 complements comparisons of MAT found in the academic literature.

The 3GPP-specified MAT considered in the report are:

\begin{itemize}
    \item \textbf{OMA} avoids inter-user interference by allocating users to orthogonal time-frequency resources.
    \item \textbf{MU-MIMO} spatially multiplexes users on the same time-frequency resources using different precoding vectors. A UE that Treats Interference as Noise (TIN) does not require network assistance information. To improve the performance, the specified MU-MIMO interference mitigation advanced receiver jointly demodulates the desired and interfering signals using network assistance information on the modulation orders of co-scheduled users. This processing is performed by a Reduced-complexity Maximum Likelihood (R-ML) receiver. The interfering signal is neither fully decoded nor cancelled.
    \item \textbf{MUST Case 1}, specified for LTE, superposes the transmissions of two users through the same precoding vector under near- and far-UE conditions. The near UE receives network assistance information on the presence and relative power of the interference, enabling it to suppress the far-UE signal before decoding its own signal. The far UE treats interference from the near UE as noise \cite{mat26}.
\end{itemize}

The candidate MAT considered in the report are:

\begin{itemize}
    \item \textbf{Power-domain NOMA}, as for MUST Case 1, exploits differences in the users' channel conditions but spatially multiplexes users over the same time-frequency resources using different precoding vectors \cite{orswieck24}. Power levels and bitrates are designed such that inter-user interference can be handled using Successive Interference Cancellation (SIC) or R-ML processing. UEs performing SIC require network assistance information on the modulation order and coding rate of the interfering signals, which they decode and cancel. With R-ML processing, interference is jointly demodulated and suppressed, but not decoded and cancelled. This NOMA variant requires the same number of DM-RS as MU-MIMO.
    \item \textbf{RSMA} also employs spatial multiplexing over shared time-frequency resources; however, it decomposes the user's messages into common and private components \cite{mao18}. The common parts are multiplexed into a single message that is decoded by all scheduled UEs. The common stream allows RSMA to achieve more flexible interference management compared to both MU-MIMO and NOMA. RSMA requires message-splitting and message-combining processing, network assistance information for decoding the common message, and an additional DM-RS for the common message compared with MU-MIMO.
    \item \textbf{CA MU-MIMO} superposes groups of MU-MIMO transmissions and uses previously cached information to cancel inter-group interference \cite{Lampiris18}. Its applicability is therefore restricted to cacheable content, while intra-group interference is handled using conventional TIN or R-ML processing \cite{mat26}.
\end{itemize}

The evaluation in GR MAT 001 assumes a simple line-of-sight channel model with varying power imbalance and channel correlation between scheduled users. Low complexity precoders are considered to align the analysis with practical implementation constraints. The results, verified through multiple independent implementations, show that candidate-MAT benefits depend on the channel conditions and optimisation objective. Power-domain NOMA can improve weak-UE bitrate relative to OMA when there is a power imbalance between scheduled users. Power-domain NOMA can also provide gains over MU-MIMO with TIN receivers when power imbalance is combined with highly correlated user channels. RSMA provides gains over MU-MIMO with TIN receivers when the scheduled-user channels are highly correlated and can provide weak-UE bitrate gains under power imbalance. CA MU-MIMO was not evaluated because suitable simplified performance metrics were unavailable.

The evaluations used capacity expressions assuming Gaussian signaling and infinite block length. Consequently, techniques employing R-ML receivers were not evaluated. The results identify promising operating conditions and implementation implications, but do not constitute definitive link-level or system-level comparisons \cite{mat26}.

\section{Ongoing Work and Relevance to 3GPP 6G}
ISG MAT is revising GR MAT 001 through work item RGR/MAT-001, considering additional evaluations, a new call for proposals, and improvements based on industry feedback. ISG MAT has informed 3GPP of both its establishment and the publication of GR MAT 001 V1.1.1 through liaison statements RP-250031 and R1-2600579, respectively. Feedback from industry, including 3GPP, is welcomed to help refine the next revision.

An important objective is to progress from the capacity-based analysis in V1.1.1 towards Link-Level Simulations (LLS) of the 5G NR Physical Downlink Shared Channel (PDSCH). These simulations use specified modulation, channel coding, and physical-layer procedures, enabling more realistic comparisons of MAT and UE receiver capabilities. In particular, the planned evaluation compares MU-MIMO with TIN receivers against MU-MIMO with R-ML advanced receivers and network assistance information, which could not be assessed using the capacity expressions in V1.1.1 \cite{mat26}.


%
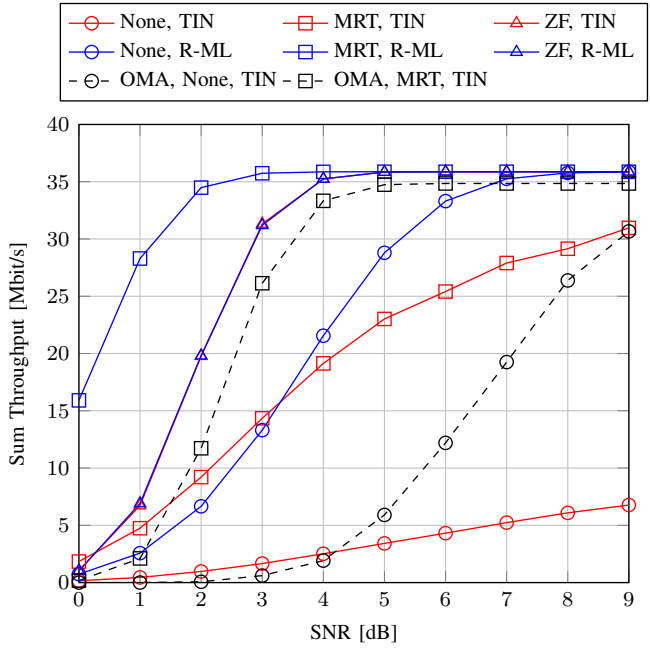
\begin{figure}
    \centering

\begin{tikzpicture}

\pgfplotsset{every axis legend/.append style={
        cells={anchor=west}, at={(0.5,1.05)}, anchor=south}}
    \pgfplotsset{every axis/.append style={line width=0.5pt}}
    \pgfplotsset{every axis/.append style={mark options=solid, mark size=2.5pt}}

\begin{axis}[
    width=\columnwidth, 
    xlabel={SNR [dB]},
    ylabel={Sum Throughput [Mbit/s]},
    grid=major,
    xmin=0, xmax=9,
    ymin=0, ymax=40,
    xtick={0,1,...,9},
    ytick={0,5,...,40},
    legend columns=3
]
\addplot+[red, solid, mark=o] coordinates {
    (0,0.1596304)
    (1,0.4484)
    (2,0.968544)
    (3,1.65908)
    (4,2.5146272)
    (5,3.4203952)
    (6,4.313608)
    (7,5.2355184)
    (8,6.0874784)
    (9,6.77084)
};
\addlegendentry{None, TIN}
\addplot+[red, solid, mark=square] coordinates {
    (0,1.8366464)
    (1,4.7458656)
    (2,9.2083424)
    (3,14.3452128)
    (4,19.137712)
    (5,23.0190624)
    (6,25.4117248)
    (7,27.9030352)
    (8,29.1513808)
    (9,30.9700912)
};
\addlegendentry{MRT, TIN}
\addplot+[red,solid,mark=triangle] coordinates {
    (0,1.0546368)
    (1,6.7295872)
    (2,19.7726464)
    (3,31.3198432)
    (4,35.2370656)
    (5,35.8468896)
    (6,35.872)
    (7,35.872)
    (8,35.872)
    (9,35.872)
};
\addlegendentry{ZF, TIN}
\addplot+[blue, solid, mark=o] coordinates {
    (0,0.735376)
    (1,2.5702288)
    (2,6.6560496)
    (3,13.3103056)
    (4,21.5536912)
    (5,28.796248)
    (6,33.3089456)
    (7,35.2550016)
    (8,35.7697648)
    (9,35.8648256)
};
\addlegendentry{None, R-ML}
\addplot+[blue, solid, mark=square] coordinates {
    (0,15.9110256)
    (1,28.2958336)
    (2,34.4855472)
    (3,35.7410672)
    (4,35.863032)
    (5,35.872)
    (6,35.872)
    (7,35.872)
    (8,35.872)
    (9,35.872)
};
\addlegendentry{MRT, R-ML}
\addplot+[blue, solid, mark=triangle] coordinates {
    (0,1.0259392)
    (1,6.9089472)
    (2,19.8174864)
    (3,31.1978784)
    (4,35.2496208)
    (5,35.845096)
    (6,35.8702064)
    (7,35.872)
    (8,35.872)
    (9,35.872)
};
\addlegendentry{ZF, R-ML}
\addplot+[black, dashed, mark=o] coordinates {
    (0,0)
    (1,0)
    (2,0.069696)
    (3,0.60984)
    (4,1.91664)
    (5,5.906736)
    (6,12.1968)
    (7,19.25352)
    (8,26.379936)
    (9,30.66624)
    (10,32.93136)
    (11,34.255584)
    (12,34.638912)
    (13,34.830576)
    (14,34.848)
};
\addlegendentry{OMA, None, TIN}
\addplot+[black, dashed, mark=square] coordinates {
    (0,0.17424)
    (1,2.108304)
    (2,11.726352)
    (3,26.136)
    (4,33.332112)
    (5,34.743456)
    (6,34.848)
    (7,34.848)
    (8,34.848)
    (9,34.848)
    (10,34.848)
    (11,34.848)
    (12,34.848)
    (13,34.848)
    (14,34.848)
};
\addlegendentry{OMA, MRT, TIN}
\end{axis}
\end{tikzpicture}

\vspace{-8pt}

    \caption{Example 5G NR PDSCH link-level evaluation comparing MU-MIMO with TIN and R-ML receivers and OMA for different precoding schemes. 4Tx2Ue2Rx, TDL-C300, 52 RBs, 30kHz SCS, 12 OFDM symbols, front-loaded DMRS (ports 0,2), perfect CSI, no HARQ, no Doppler, Type I Single-Panel Codebook, per-RE precoding MCS 9 (MU-MIMO), MCS 15 (OMA).}
    \label{fig:lls}

\vspace{-12.5pt}
    
\end{figure}

Figure \ref{fig:lls} shows LLS results without precoding `None', Maximum Ratio Transmission `MRT', where each UE selects the precoding vector maximizing the effective channel gain, and Zero Forzing `ZF'. As expected, the evaluations show that R-ML receivers with network assistance information achieves a significant gain compared to TIN (except for ZF where all interference is suppressed at the transmitter). MU-MIMO with MRT outperforms ZF because it does not affect the power of the precoded transmit symbols. To achieve a similar sum throughput than MU-MIMO, OMA uses 26 non-overlapping RBs per UE and Modulation and Coding Scheme (MCS) 15. MU-MIMO significantly outperforms OMA \textit{solely} for R-ML receivers with network assistance information, highlighting the fact that MU-MIMO requires efficient interference mitigation schemes.  

ISG MAT has also initiated work in two additional areas. DGR/MAT-002 studies low-complexity transceiver architectures for high-order Single-User MIMO, with particular relevance to Fixed Wireless Access deployments. DGR/MAT-003 studies MAT for increasing spectral efficiency in Non-Terrestrial Networks (NTN), including inter-beam interference management, SU-MIMO and MU-MIMO, mixed unicast-multicast and multi-group multicast transmission, and NTN-specific channel characteristics and impairments.

These activities are timely in the context of the 3GPP Release 20 Study on 6G Radio, in which increased spectral efficiency is a key objective \cite{3gpp25}. The interference-mitigation studies in GR MAT 001 and its ongoing revision build on previous 3GPP studies of Network-Assisted Interference Cancellation and Suppression (NAIC) for LTE in TR 36.866, downlink MUST in TR 36.859, and NR demodulation-performance evolution in TR 38.878. The latter resulted in performance requirements in TS 38.101-4 for a MU-MIMO interference-mitigation advanced receiver that uses network assistance information on the modulation orders of co-scheduled users to support R-ML processing.

Within the 6G study, a multi-operator contribution proposed prioritising enhanced NAIC for MU-MIMO, including the provision to co-scheduled users of detailed interference characteristics such as resource allocation, modulation and coding, and DM-RS configuration \cite{bov25}. A separate multi-company contribution proposed common Downlink Control Information (DCI) carrying scheduling information for a group of MU-MIMO UEs, potentially enabling interference-mitigation techniques beyond those specified for 5G NR \cite{oda26}. These contributions demonstrate interest from a broad set of stakeholders but do not represent agreed 3GPP outcomes. By establishing common terminology, comparing candidate MAT against specified 3GPP features, and progressing towards realistic evaluations, ISG MAT can accelerate the transition of research results into technically grounded contributions to 6G standardisation discussions.

\section{Conclusions}
Improved spectral efficiency and inter-user interference mitigation are important aspects of IMT-2030 and the ongoing 3GPP 6G study. ETSI ISG MAT provides an open research and pre-standardisation setting that brings together industry, academia, research organisations, and government administrations to help close the gap between research on MAT and its consideration in 6G standardisation. GR MAT 001 contributes a standards-oriented comparison of 3GPP-specified and candidate MAT, identifying the channel conditions under which candidate techniques can provide substantial gains, as well as deriving their implications for transmitter and receiver processing, network assistance information, and DM-RS requirements. Ongoing work includes the development of realistic 5G NR link-level evaluations and the study of MAT for NTN systems. These activities can provide timely technical evidence to inform 3GPP 6G standardisation discussions.

\bibliographystyle{ieeetr}
\bibliography{ref_papers}

\end{document}